\documentclass[a4paper,fleqn,usenatbib]{mnras}
\usepackage[dvipsnames,usenames]{color}
\usepackage{colortbl}
\usepackage{mathptmx}
\usepackage{etoolbox}
\usepackage{graphicx}
\usepackage{amsmath}
\usepackage{natbib}

\newcommand\rd{{\rm d}}

\pubyear{2026}
\begin{document}
\title[Self-consistent radiative losses]{A simulation-based analytic model of radio galaxies II:
  self-consistent radiative losses}
\author[M.J.\ Hardcastle]
{M.J.\ Hardcastle\\
Centre for Astrophysics Research, Department of Physics, Astronomy and Mathematics, University of
  Hertfordshire, College Lane, Hatfield AL10 9AB, UK\\
}
\maketitle
\begin{abstract}
The evolution of the radio properties of high-redshift radio-luminous
active galactic nuclei is well known to be strongly affected by
inverse-Compton losses which increase rapidly at higher redshifts due
to the higher energy density in the cosmic microwave background
radiation. Dynamical models of these sources, however, generally
neglect the effects of radiative losses on the dynamics and energetics
of the sources themselves. In the framework of an analytical model I
developed in a previous paper, I show that the assumption that these
losses can be neglected becomes unsafe at high redshifts. The effects
on the source dynamics and energetics can
result in significantly lower predicted luminosities for high-redshift
sources in both radio (synchrotron) and X-ray (inverse-Compton) bands.
Modelling of the population of these powerful sources needs to take
account of these results in order to infer jet powers at high
redshift, and also to make a correct prediction of the number of
sources that may be available to provide a background for studies of
the 21-cm forest.
\end{abstract}

\section{Introduction}
\label{sec:intro}

Observations with current and next-generation radio telescopes
increasingly probe the active galaxy population in the distant
universe, and evidence for accreting black holes in the very early
universe is becoming more and more compelling
\citep[e.g.][]{Maiolino+24}. At least some of these accreting black
holes might be expected to fulfil the conditions required to generate
powerful jets (which are likely to be related to the spin of the black
hole and the magnetization of the material in the accretion flow:
\citealt{Blandford+Znajek77}).

In the local Universe, the effects of these jets are
predominantly detected as a result of the radio emission due to
synchrotron radiation from high-energy particles accelerated within
the jet, but the normalization of this synchrotron emission and its
spectrum depends on the details of the acceleration and evolution of
the particles within the jets and the large-scale lobes that they
inflate. It is therefore important to have a model
framework within which we can understand the expected evolution of
radio-luminous AGN (RLAGN) at high redshift. In principle this needs
to take account of (i) the evolution of the conditions required for
jet generation in the first place, (ii) the different local
environments experienced by the jets at high redshift, and (iii) the
rapid increase in the energy density of the cosmic microwave
background radiation, which goes as $(1+z)^4$ and drives
inverse-Compton losses \citep[e.g.][]{Wu+17}.

An important application of such a model would be a prediction of the
radio properties of the high-redshift radio sources that could be used
to give a continuum background against which the so-called `21-cm
forest' could be detected \citep[e.g.][]{Mellema+13}. In principle,
this absorption signal from the redshifted neutral hydrogen in the
Universe along the line of sight to the radio source would give a view
of the evolution of neutral hydrogen and particularly the timing and
evolution of the epoch of reionization which is very complementary to
studies of neutral hydrogen in emission. The study of high-redshift
neutral hydrogen provides one of the key science cases for the
forthcoming Square Kilometre Array \citep{Koopmans+15}.

In an earlier paper \citep[][hereafter Paper 1]{Hardcastle18} I
described a simulation-based analytic model that is primarily designed
to model the evolution of powerful RLAGN, where in this context
`powerful' means that the sources have enough kinetic power to drive a
shock through their external environment throughout their
evolution\footnote{This corresponds roughly to sources above the
classical Fanaroff-Riley break \citep{Fanaroff+Riley74} which implies
jet powers above around $10^{38}$ W, but of course the ability to
drive a shock is also dependent on environment.}. The key feature of
the model was that the energy supplied by the initially light,
electron-positron jet is partitioned in a predictable way between the
`lobes', containing the synchrotron-emitting particles, and the
`shocked shell', consisting of shocked and swept up baryonic matter
from the environment. Equally importantly, though, it was assumed that
energy was conserved inside the shocked shell: that is, radiative
losses from any component of the system were negligible. I showed in
Paper 1 that this assumption was reasonable for low-redshift RLAGN,
which was the principal use case for the model at the time. However,
as I noted in that paper, it seemed likely to break down badly for
high-redshift sources where inverse-Compton losses become more and
more important. In the present paper I extend the framework of Paper 1
to develop a model in which the losses and the radio source dynamics
can be considered consistently, and show that such a model has
significant observational consequences at high redshift, affecting the
inference of jet powers, the population of high-$z$ luminous radio
sources which it is hoped may provide a direct detection of the 21-cm
forest, and predictions for future sensitive X-ray surveys.

Throughout the paper I use a standard flat $\Lambda$CDM cosmology with
$H_0 = 70$ km s$^{-1}$, $\Omega_{\rm m} = 0.3$ and $\Omega_\Lambda =
0.7$, and define spectral index $\alpha$ in the sense $S \propto
\nu^{-\alpha}$.

\section{Losses}

\subsection{The problem}
\label{sec:problem}

Let us start by revisiting some of the assumptions of Paper 1. We
describe a RLAGN as having a two-sided jet power $Q$, and since the
jet is initially light and relativistic this implies a jet momentum
flux per lobe of $Q/2c$. As noted above, in the absence of losses, all
of the jet power goes into the lobes and the shocked shell (the work
done on the external environment, in this phase, is all done on the
shocked material -- by construction, the unshocked medium is
unaffected by the AGN). Therefore the total energy in the lobes and
shocked shell at any given time $t$ is just $Qt$. The key simplifying
assumption, which is supported by simulations, is that a constant
fraction $\xi$ of the energy supplied by the jet is stored in the
lobes, while the rest is in the shocked shell. In Paper 1 I went on to
derive a dynamical model for the expansion of the lobes through the
external medium, which is unchanged in the present paper.

However, it is easy to see that at some point a model like this will
lead to an inconsistency. Consider inverse-Compton losses, which for
a single electron\footnote{Throughout the paper, except where
otherwise stated, I assume that the
energy density of the lobes is dominated by a bulk neutral
electron/positron plasma; `electron' should be taken to refer to
electrons of both charges.} of Lorentz factor $\gamma$ ($\gamma \gg 1$) and a
radiation field of energy density $U_{\rm rad}$ can be written \citep{Longair10}
\begin{equation}
  \frac{\rd E}{\rd t} = -\frac{4}{3} \sigma_T c U_{\rm rad} \gamma^2
  \label{eq:loss}
\end{equation}
where $\sigma_T$ is the Thomson cross-section and $c$ is the speed of
light. For the important case of inverse-Compton scattering from the
CMB, we know that $U_{\rm rad} = 4.2\times 10^{-14}(1+z)^4$ J
m$^{-3}$. The strong dependence on redshift means that the energy
density in the CMB, and therefore the loss rate for an electron of a
given energy, is 2,400 times higher at $z=6$ than at $z=0$.

Suppose that the electron energy distribution in the lobes is given by
a power law,
\begin{equation}
  N(E) = N_0 E^{-q}
\end{equation}
where $N(E)$ here is the differential number density of electrons of a
particular energy $E$ and $N_0$ is a normalizing factor. Then we have,
assuming as a slight simplification that the energy density in electrons is dominant in the
lobes, that
\begin{equation}
  \xi Qt = V_{\rm lobe} \int_{E_{\rm min}}^{E_{\rm max}} E N(E) \rd E = V_{\rm lobe}N_0 \int
_{E_{\rm min}}^{E_{\rm max}}  E^{1-q} \rd E  = V_{\rm lobe} N_0 I_{\rm
    E}
\label{eq:energy}
\end{equation}
with
\begin{equation}
I_{\rm E} = \left\{ \begin{array}{ll}\ln(E_{\rm max}/E_{\rm min})&q=2\\
{\frac{1}{2-q}} \left[E^{(2-q)}_{\rm max}-E^{(2-q)}_{\rm min}\right]&q\neq
2\\
\end{array}\right .
\end{equation}

Now the inverse-Compton loss rate for this particle population is
given by integrating Eq.\ \ref{eq:loss} over the energies of all the
available particles:
\begin{equation}
-\frac{\rd E_{\rm total}}{\rd t} = \frac{4}{3} V_{\rm lobe}
\sigma_T c U_{\rm rad} \int_{E_{\rm min}}^{E_{\rm max}} N(E)
\left(\frac{E}{m_e c^2}\right)^2 \rd E
\end{equation}
giving
\begin{equation}
  -\frac{\rd E_{\rm total}}{\rd t} = \frac{4}{3} V_{\rm lobe} N_0
  \frac{\sigma_T U_{\rm rad}}{m_e^2 c^3} I_{\rm loss}
\end{equation}
where similarly
\begin{equation}
I_{\rm loss} = \left\{ \begin{array}{ll}\ln(E_{\rm max}/E_{\rm min})&q=3\\
{\frac{1}{3-q}} \left[E^{(3-q)}_{\rm max}-E^{(3-q)}_{\rm min}\right]&q\neq
3\\
\end{array}\right .
\end{equation}
The assumption of negligible radiative losses
breaks down if the radiative loss becomes comparable to the input jet
power, i.e., if $Q \approx \left|\rd E/\rd t\right|$. Writing down
this condition in the case of inverse-Compton losses as an inequality gives
\begin{equation}
  Q < \frac{4}{3} V_{\rm lobe} N_0
  \frac{\sigma_T U_{\rm rad}}{m_e^2 c^3} I_{\rm loss}
\end{equation}
for the assumption to break down. But we can use Eq.\ \ref{eq:energy} to eliminate $V_{\rm lobe}N_0$
from this equation:
\begin{equation}
  Q < \frac{4}{3} \frac{\xi Qt}{I_{\rm E}}  \frac{\sigma_T U_{\rm
      rad}}{m_e^2 c^3} I_{\rm loss}
\end{equation}
Hence these assumptions must have broken down, \textit{independent of $Q$}, by
a time $t$ given by
\begin{equation}
  t = \frac{3}{4} \frac{m_e^2 c^3}{\xi \sigma_T U_{\rm rad}}
  \frac{I_{\rm E}}{I_{\rm loss}}
\end{equation}
We see that the timescale $t$ on which our approximation breaks down
is inversely proportional to the energy density of the CMB, and thus
goes as $(1+z)^{-4}$. $t$ is
somewhat sensitive to the parameters of the power-law distribution,
particularly $q$ and $E_{\rm max}$, through the ratio $I_{\rm
  E}/I_{\rm loss}$, but adopting $q=2.1$ (as used in Paper 1) and
$E_{\rm max}$ corresponding to $\gamma = 10^4$ (it cannot be less than
this if we see radiating electrons at GHz frequencies), and
substituting the energy density of the CMB for $U_{\rm rad}$, we find
a critical timescale of over 5 Gyr at $z=0$, but only around 2 Myr at
$z=6$. Thus, while the effects of inverse-Compton losses can be safely
neglected for low-redshift RLAGN, whose lifetimes are likely to be
mostly $\ll 1$ Gyr, they absolutely cannot be neglected
for objects at high redshift, since Myr timescales are well within the
range of lifetimes expected for RLAGN. Indeed, the longest-lived
objects should already start to be affected by $z=1$. Crucially, in this
approximation all RLAGN will be affected in the same way independent
of their jet power, because to first order inverse-Compton losses
scale linearly with the total integrated energy in the lobes, as noted
in Paper 1.

Synchrotron losses are more complex because the overall synchrotron
loss history depends on the evolving magnetic field strength in the
lobes. The loss rate to synchrotron emission is the same as in
eq.\ \ref{eq:loss} but with the magnetic field energy density, $U_B$,
being substituted for the radiation energy density. If we assume (as
in Paper 1) that the magnetic field total energy is a fixed fraction
$\zeta$
of the energy in radiating particles, then for $\zeta \ll 1$ we have
\begin{equation}
  U_B = \zeta N_0 I_{\rm E} = \frac{\zeta\xi Qt}{V_{\rm lobe}}
\end{equation}
Then our assumptions become invalid when
\begin{equation}
  Q < \frac{4}{3} \frac{\xi Qt}{I_{\rm E}} \frac{\sigma_T}{m_e^2c^3}
  \frac{\zeta\xi Qt}{V_{\rm lobe}} I_{\rm loss}
\end{equation}
or, rearranging,
\begin{equation}
  \frac{Qt^2}{V_{\rm lobe}} > \frac{3}{4}\frac{m_e^2c^3}{\sigma_T}
    \frac{1}{\zeta \xi^2}\frac{I_{\rm E}}{I_{\rm loss}}
\end{equation}
where the right-hand side contains only constants of the model. Unlike
the inverse-Compton case, this depends on the dynamics through the
time-dependent $V_{\rm lobe}$. In the very early parts of the model of
Paper 1, $V_{\rm lobe} \approx c^3t^3$, and so the synchrotron loss
will always exceed the jet power as $t \to 0$. This feature of the
model can be ignored, as it is only relevant to the time period when
the jet has just turned on for the first time, and we can reasonably
assume that e.g. particle acceleration or the mechanism to establish
local quasi-equipartition of energy do not operate effectively at
these times. Otherwise we expect to see any problems only for high $Q$
and at late times if $V_{\rm lobe}$ grows more slowly than $t^2$. In
practice, this does not occur in even the most extreme modelled jets
(Paper 1) and so catastrophic synchrotron losses can safely be
neglected.

Finally, I noted in Paper 1 that thermal bremsstrahlung losses from
the shocked shell appear likely to be negligible in comparison to
those from the two non-thermal processes. Since these are also
directly dependent on the external environment, I do not attempt to provide an
analytic description of them here.

\subsection{Incorporating losses self-consistently in the dynamical model}

The implementation of bolometric radiative losses in the original code
of Paper 1 was already more sophisticated than the simple calculation
described above, because it models the electron energy spectrum of the
lobes as a sum of discrete \cite{Jaffe+Perola73} aged spectra, as
described in section 2.3 of Paper 1, and takes account of adiabatic
losses while doing so. Thus it models the true time evolution of the
term ${I_{\rm E}}/{I_{\rm loss}}$ in the analysis above. It also does
not neglect the contribution of the lobe magnetic fields to the
energetics. However, it is difficult to modify the dynamical solver of
the code, which essentially solves the coupled differential equations
for lobe expansion in the longitudinal and transverse directions, to
include these realistic losses: for example, the aged spectra depend
on the time history of the magnetic field strength which is not
readily available inside the numerical solver. Equally, however, the
losses cannot just be calculated outside the dynamical solver as was
done for Paper 1, because they affect the dynamics through the lobe
pressure.

To overcome this problem I implemented an iterative solution. From an
existing run of the code, the losses can be calculated, and therefore
the total energy in the system taking account of radiative losses,
\begin{equation}
E(t) = \int_0^t (Q - L_{\rm IC} - L_{\rm synch} - L_{\rm Brems}) \rd t'
\end{equation}
can be written down, where the three loss terms are respectively due
to inverse-Compton, synchrotron and thermal bremsstrahlung from the
shocked shell. $E(t)$ is the quantity that affects the dynamics,
and which is approximated by $Qt$ in a typical run of the code. Having
generated a table of $E(t)$, the code can be run again with the same
input parameters but with a linear interpolation of $E(t)$ substituted
for $Qt$ in the dynamical and loss calculation parts of the code. (The
momentum flux of the jet is unchanged as the radiation processes do
not carry away significant net momentum.) This approach converges quite
quickly on a solution if $E(t)$ remains positive throughout, but if it
does not (i.e. the inverse-Compton emission would remove more energy
from the system than it actually contains) then the lookup table for
$E(t)$ must be truncated in time, and then more iterations are
necessary for convergence. In these cases the iterative code
alternates between solving for the dynamics given the losses for the
previous iteration, with no extrapolation, and solving for the losses
assuming a continued loss rate after the end of the previous iteration
which is equal to the time-weighted average loss rate over the source
lifetime. The iterative code can then be run until the typical loss
rate converges. In the implementation used for this paper, we require
the lifetime of the run not to be increasing and to have converged
within 2 per cent, and the relative mean absolute difference between
loss rates in the past ten timesteps to be within 10 per cent, in
order to decide that the iterations have converged.

\section{Results}

\begin{figure*}
  \includegraphics[width=\linewidth]{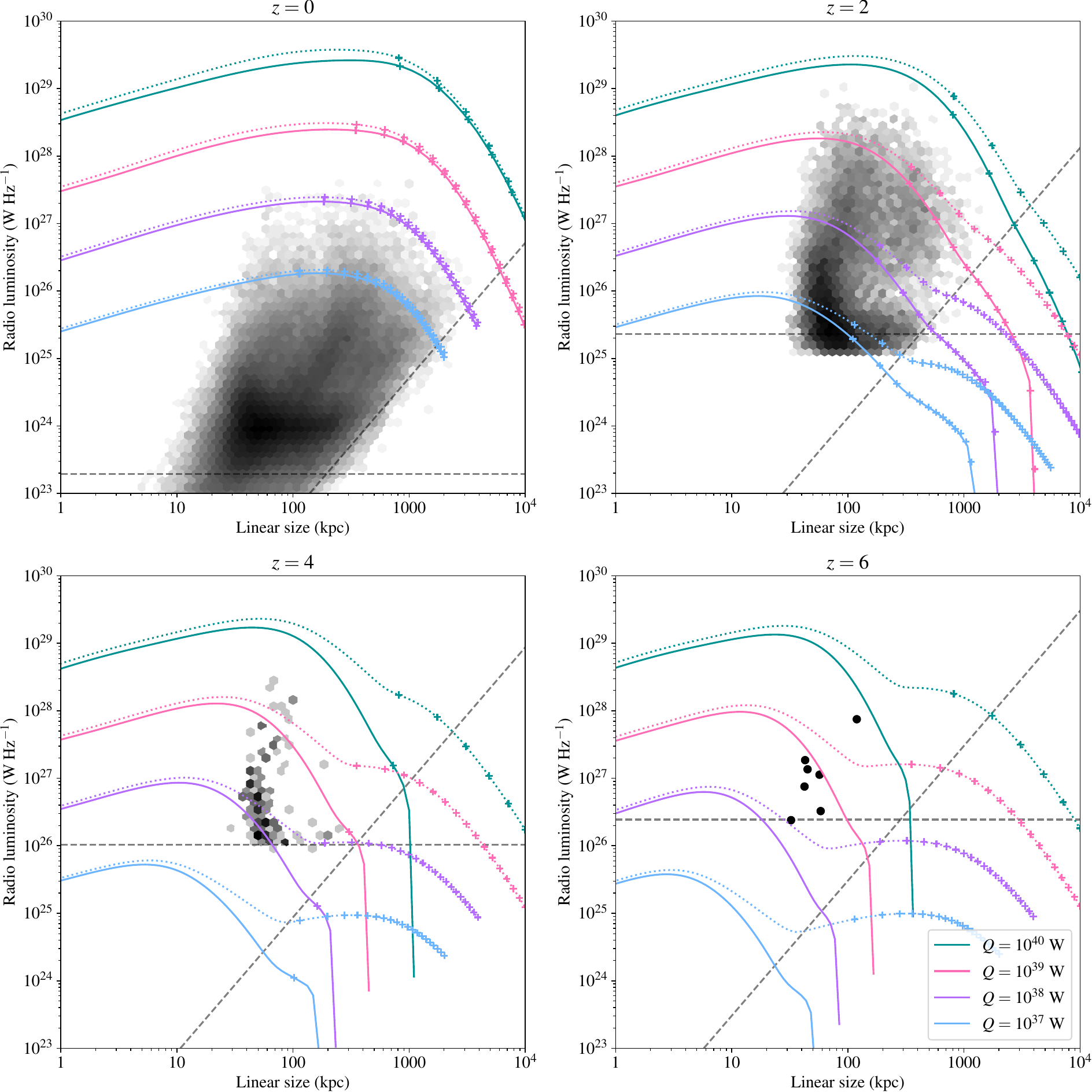}
  \caption{Evolution of radio sources in the power-linear size diagram
    for different jet powers and redshifts. Radio luminosity is
    measured at 144 MHz: the rest-frame luminosity at $144(1+z)$ MHz
    is calculated and then $k$-corrected back to 144 MHz assuming
    $\alpha=0.7$ to mimic the approach used in observations. Panels
    show the same jet powers and environmental conditions for $z=0$
    (top left), $z=2$ (top right), $z=4$ (bottom left) and $z=6$
    (bottom right). In all cases four jet powers ($Q=10^{37}$ W,
    $Q=10^{38}$ W, $Q=10^{39}$ W, $Q=10^{40}$ W) are represented by
    coloured lines. Dotted lines represent the evolution in the
    reference model of Paper 1 and solid lines show the new
    self-consistent model. Crosses on both sets of lines indicate
    50-Myr intervals of the source's evolution: lines without a cross
    terminate before the source reaches an age of 50 Myr. Black
    density plots, or in the case of $z=6$ black individual points,
    represent the positions of LOFAR RLAGN at these redshifts from
    \protect\cite{Hardcastle+25}, where we include all AGN in that
    catalogue with redshift within $\pm 0.5$ of the central redshift.
    As discussed by \protect\cite{Hardcastle+23}, the $z\approx 6$
    points in this catalogue are all taken from
    \protect\cite{Gloudemans+22}. Note that many of the sources on
    this plot are unresolved, and so their linear sizes as plotted
    should be taken as upper limits on the true projected linear size.
    In particular none of the $z\approx 6$ sources is resolved. Grey
    dashed lines represent the flux limit (horizontal line) and
    surface brightness limit (diagonal line) of the LOFAR-based
    \protect\cite{Hardcastle+25} catalogue at the central redshift of
    the sources plotted.}
    \label{fig:pd}
    \end{figure*}

To see the additional effect of self-consistent losses we need to run
the code with and without the iterative loss finding approach
described above. The code already accounts for radiative and adiabatic
losses in the integrated synchrotron spectrum, and therefore the
easiest way to represent the different models is as tracks in the
power-linear size ($PD$) diagram. I ran the code for four redshifts
($z=0$, 2, 4, 6) and four jet powers ($Q = 10^{37}$, $10^{38}$,
$10^{39}$ and $10^{40}$ W), modelling evolution over a maximum of 1
Gyr. The highest jet power here is of order the Eddington luminosity
for a $10^9$ solar mass black hole, and so represents something like
the maximum jet power we would expect to observe in the Universe,
comparable to that of a powerful source like Cygnus A at $z=0$,
while $Q = 10^{37}$ W is the power observed for the top end of the
Fanaroff-Riley class I (FRI) population at $z=0$; thus these models
span the full range of powerful RLAGN. For simplicity all models
use the same environment, a universal pressure profile
\citep{Arnaud+10} environment with $M_{500} = 2.5 \times 10^{13}
M_\odot$, which is not unrealistic for low-redshift radio galaxies; a caveat (which I return to below) is that RLAGN environments are likely to be
significantly different at high redshifts, but we keep the
environments constant here in order to see the purely
redshift-dependent effects of the model.

Results for the luminosity and size of the modelled sources at
rest-frame 144 MHz are plotted as a function of redshift in
Fig.\ \ref{fig:pd}, where I also show the results of the original
model without the iterative search for a self-consistent solution and
the distribution of AGN at those redshifts from \cite{Hardcastle+25}
on the $PD$ diagram. (The \cite{Hardcastle+25} AGN catalogue was based
on the optical identification catalogue of \cite{Hardcastle+23}, who
took all $z>5$ objects from the work of \cite{Gloudemans+22} on
high-redshift LoTSS quasars, and so these are the objects plotted for
the $z=6$ redshift bin.)

Several features of this figure are of interest. Firstly, we see that,
as expected, the self-consistent model gives lower radio luminosity than the
Paper 1 model for a given linear size for all redshifts, with the
difference increasing with redshift. The offset at $z=0$ is a result
of synchrotron losses at early times, but otherwise at $z=0$ we see
almost no effect of the self-consistent modelling, as the late-time
inverse-Compton losses are negligible. At higher redshifts we see an
increasing difference between the Paper 1 models and the
self-consistent ones with source size/age, corresponding to the
increasing importance of inverse-Compton losses with time. Crucially,
in the bottom panels of Fig.\ \ref{fig:pd}, the tracks in the $PD$
diagram decline steeply once the sources reach physical sizes of
10-100 kpc, as there is no way for them to avoid catastrophic
inverse-Compton losses. Relative to the Paper 1 models, there is a
large region of parameter space in the $PD$ diagram at large physical
sizes that is excluded for sources below a given jet power $Q$. At
high $z$, sources start their steep downward trajectory in radio
luminosity after only a few tens of Myr in age. The differences
between the predicted radio luminosities with and without taking
account of self-consistent radiative losses can be as much as two
orders of magnitude at the point where the simulated sources intercept
the observational surface brightness limit.

\begin{figure*}
  \includegraphics[width=\linewidth]{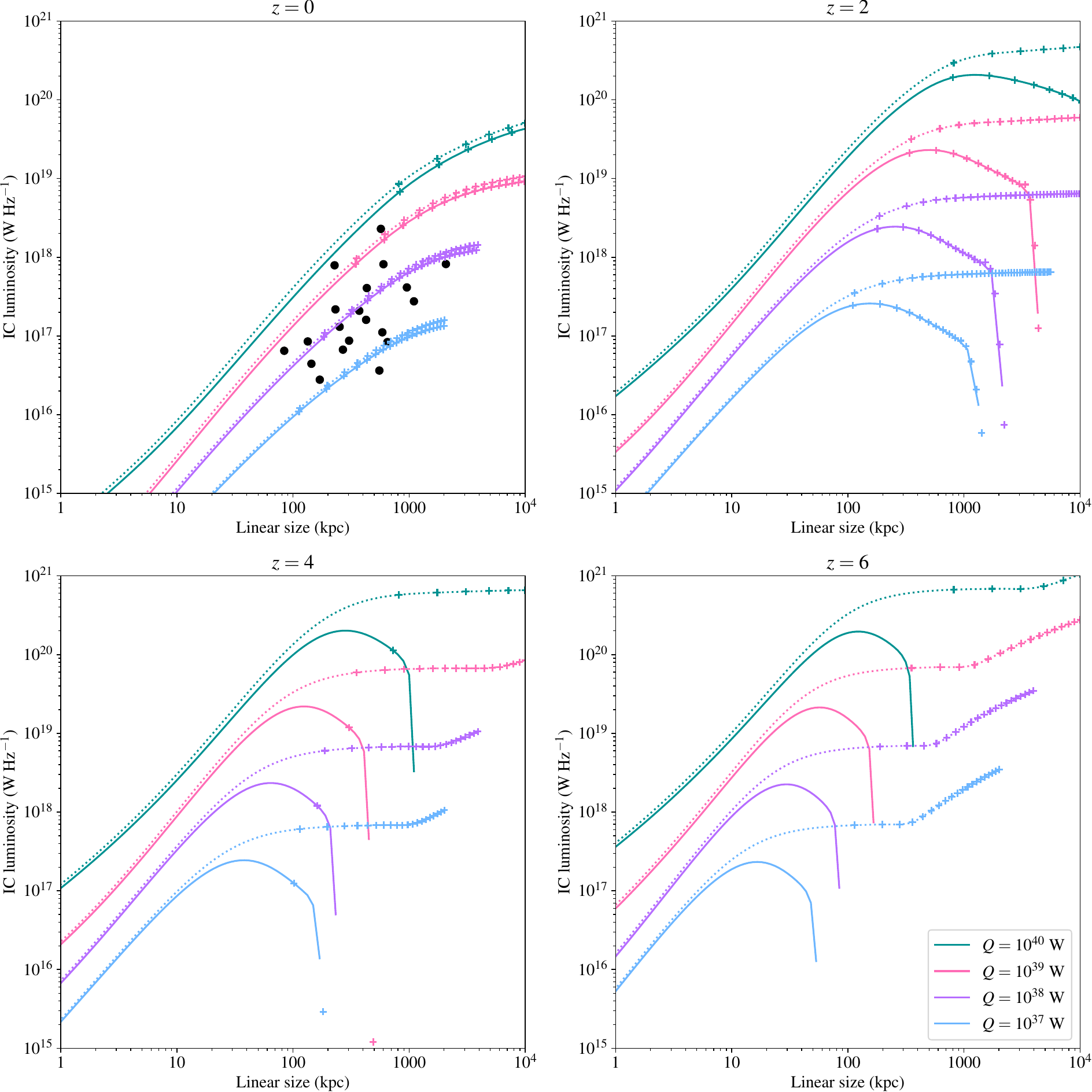}
  \caption{Evolution of radio sources in the inverse-Compton power-linear size diagram
    for different jet powers and redshifts. As Fig.\ 1, but here we
    plot the inverse-Compton luminosity at a rest-frame energy of 1
    keV ($2.4 \times 10^{17}$ Hz). Black data points in the $z=0$
    panel are the $z<0.25$ objects from \protect\cite{Ineson+17}.}
    \label{fig:icpd}
\end{figure*}

Another interesting effect is on the evolution of inverse-Compton
emission itself, and this is shown in Fig.\ \ref{fig:icpd}. The
inverse-Compton $PD$ diagram plotted in the Figure is not commonly
used observationally because sample sizes of sources with good
inverse-Compton measurements are rare, but it can be seen that the
model predictions for low-redshift sources are very consistent with
observation. However, the predictions for high-redshift sources are
very different once the self-consistent losses are included. The
maximum achievable inverse-Compton luminosity is greatly reduced at
high $z$ and in fact is more or less independent of redshift for a
given jet power in the self-consistent models.

Other predictions of the self-consistent models are similar to those
of the Paper 1 model. In particular, the evolution of spectral index
is very similar, and sources grow to roughly the same size as a
function of time, because the longitudinal expansion of the lobes is
driven by the jet momentum flux, which is unaffected by radiative
losses. The predicted axial ratios of sources in the self-consistent
models are systematically smaller at higher redshift, as expected
since the losses remove pressure from the lobes and hence reduce the
rate at which they expand transversely. This might tend to reduce
  the feedback effects of these high-redshift sources.

Of course, the use of a single constant environment over the wide
range of redshift (and even jet power) modelled here is completely
unrealistic, and is intended only to allow the reader to isolate the
effects of redshift-dependent losses from all other effects. All other
things being equal, we expect typical radio galaxy environments to be
poorer (lower $M_{500}$) in the past, but also to potentially have a
lower ratio of hot baryonic mass to dark matter mass and a flatter
radial pressure distribution. Lower densities of hot gas tend to make
lobes expand faster and so to be less luminous for a given size, but
flatter hot gas distributions have the opposite effect
\citep{Hardcastle+Krause13} and so might push the onset of
catastrophic inverse-Compton losses to sources of smaller sizes. All
of this, however, really only affects the size-luminosity
relationship: the timescale on which inverse-Compton losses start to
affect the dynamics is essentially unaffected by environment, by the
arguments of Section \ref{sec:problem}.

\section{Discussion and conclusion}

In the previous sections I have shown the need for self-consistent
modelling of the effects of radiative losses on the dynamical
modelling of radio sources from Paper 1, and implemented a simple
scheme for making the models self-consistent in the situation where
radiative losses are not negligible in terms of the overall energetics
of the system. Results are shown in Figs \ref{fig:pd} and \ref{fig:icpd}.

Interestingly, the observations shown in Fig.\ \ref{fig:pd} are at
least qualitatively consistent with the models described in the
previous section. At low redshifts, observed source linear sizes are
limited not by RLAGN physics but by the surface brightness limit in
the LOFAR observations (which either causes the sources not to be
observed at all or to have underestimated sizes) and so the $PD$
diagram has a sharp boundary at its right-hand edge, as seen in the
top two panels of Fig.\ \ref{fig:pd}. But at high $z$ the observed
LOFAR sources are all compact and most populate the area below the $Q
= 10^{39}$ W line on the plots -- the observational selection effects
alone would clearly (from Fig.\ \ref{fig:pd}) allow a population of
larger sources to exist, but they do not. Although this could be for
other reasons (such as the inadequacy of a single environmental model
over all of cosmic time, or a source lifetime distribution at high $z$
which does not allow sources to grow to large sizes), and it is worth
noting that the high-$z$ sources are typically quasars and are likely
to appear somewhat smaller due to projection, it is a general
prediction of the model of this paper that large, luminous sources
will be very rare and that the most luminous objects ($\ga 10^{28}$ W
Hz$^{-1}$) will only remain in that state for a short period. These
effects would of course be even more extreme for redshifts $z>4$.

These results have implications for the inference of jet powers from
radio observations, such as has been carried out using the Paper 1 models
by \cite{Hardcastle+19} and \cite{Pierce+26a}. The effects of
self-consistent losses can safely be ignored at low $z$ but by $z
\approx 2$ the use of the Paper 1 model (or any other which neglects
radiative losses in the source dynamics) would give rise to an
underestimate of the jet power which could be quite significant, since
it scales close to linearly with the difference in low-frequency
luminosity between the two models. This would have the effect of
pushing up our estimates of integrated kinetic power as a function of
redshift. We will attempt to account for this effect in future analysis.

The results may also have implications for the prospects for studying
the 21-cm forest using luminous RLAGN targets to give a direct
detection of absorption lines. At the
rest-frame GHz frequencies of the required observations, powerful
RLAGN at low $z$ are mostly dominated by radio emission from the
lobes, which is what is modelled by the semi-analytical models
discussed here and in Paper 1. Only a small minority of all RLAGN,
aligned close to our line of sight, have observed radio luminosity
dominated by the relativistically beamed core and jet. Thus, in
considering the population of objects that might provide a bright
background for 21-cm studies, it is important to take account of the
physics outlined here and in Paper 1 rather than assuming some scaling
from the optical properties of quasars \cite[cf., e.g.,][]{Niu+25}. Sources with jet powers in the
range $10^{39}$ -- $10^{40}$ W, which are required to produce 144-MHz radio
flux densities from the lobes of tens of mJy at high redshift, will
maintain the required radio luminosity for only a few to a few tens of
Myr and so will be substantially rarer than their counterparts at low
$z$, though of course this does not rule out their existence
altogether \citep[e.g][]{Gloudemans+25}.

Finally, the models predict (Fig.\ \ref{fig:icpd}) that
inverse-Compton luminosities of powerful sources essentially have a
jet-power-linked maximum at high $z$, despite the strong increase in
the energy density of the CMB. This would affect the conclusions of
e.g. \cite{Mocz+11} and make inverse-Compton `ghosts' less prevalent
at high redshift, in turn affecting the expectations for observations
with next-generation X-ray instruments such as \textit{New-Athena}.

Detailed predictions of the population of RLAGN at these redshifts
should take account of the jet generation mechanism, including the
distribution of black hole mass, spin and accretion rate, as well as
the expected small-scale and large-scale enviroments which, together
with the jet power, determine the source dynamics. Clearly such an
effort is beyond the scope of this paper, but forward-modelling of the
radio AGN population from cosmological simulations is starting to
become realistic: such efforts are complicated by, but very clearly
need to take into account, the loss processes that I have discussed
above.

\section*{Acknowledgments}

I acknowledge support from the UK Science and Technology Facilities
Council [ST/Y001249/1], and am grateful to Judith Croston and Jonathon
Pierce for comments on the manuscript. I also thank the referee,
Patrick Yates-Jones, for helpful comments on the submitted version of
the paper. This research made use of {\sc Astropy}, a
community-developed core Python package for astronomy
\citep{AstropyCollaboration13} hosted at \url{http://www.astropy.org/}
and of {\sc Matplotlib} \citep{Hunter07}.

LOFAR is the Low Frequency Array designed and constructed by ASTRON. It has observing, data processing, and data storage facilities in several countries, which are owned by various parties (each with their own funding sources), and which are collectively operated by the LOFAR ERIC under a joint scientific policy. The LOFAR resources have benefited from the following recent major funding sources: CNRS-INSU, Observatoire de Paris and Université d'Orléans, France; BMBF, MIWF-NRW, MPG, Germany; Science Foundation Ireland (SFI), Department of Business, Enterprise and Innovation (DBEI), Ireland; NWO, The Netherlands; The Science and Technology Facilities Council, UK; Ministry of Science and Higher Education, Poland; The Istituto Nazionale di Astrofisica (INAF), Italy.

This research made use of the Dutch national e-infrastructure with support of the SURF Cooperative (e-infra 180169) and the LOFAR e-infra group. The Jülich LOFAR Long Term Archive and the German LOFAR network are both coordinated and operated by the Jülich Supercomputing Centre (JSC), and computing resources on the supercomputer JUWELS at JSC were provided by the Gauss Centre for Supercomputing e.V. (grant CHTB00) through the John von Neumann Institute for Computing (NIC).

This research made use of the University of Hertfordshire high-performance computing facility and the LOFAR-UK computing facility located at the University of Hertfordshire and supported by STFC [ST/P000096/1], and of the Italian LOFAR IT computing infrastructure supported and operated by INAF, and by the Physics Department of Turin university (under an agreement with Consorzio Interuniversitario per la Fisica Spaziale) at the C3S Supercomputing Centre, Italy.

\section*{Data availability}

No data were newly processed for this paper. Catalogued data for the LOFAR AGN
sample used in the plots is available at
\url{https://lofar-surveys.org/dr2_release.html}.

\bibliographystyle{mnras}
\renewcommand{\refname}{REFERENCES}
\bibliography{mjh,cards}

\end{document}